\documentclass{PoS}
\usepackage{setspace}
\def\Msun{{\rm\,M_\odot}}
\def\whz {{\rm\, W \,Hz^{-1}}}

\newcommand{\angstrom}{\mbox{\normalfont\AA}}
\title{Radio Properties of Narrow-Line Seyfert 1 Galaxies}

\ShortTitle{Radio Properties of NLSy1s}

\author{\speaker{Matthew L. Lister}\\
        Department of Physics and Astronomy, Purdue University, USA\\
        E-mail: \email{mlister@purdue.edu}}

      \abstract{ The last decade has witnessed a steadily increasing
        number of observational studies concerning the rare class of
        radio loud narrow-line Seyfert 1 galaxies, of which several
        hundred are currently known. According to the current AGN
        paradigm, the low black hole masses and high accretion rates
        of narrow-line Seyfert 1 galaxies (NLSy1) should make them
        unlikely to launch jets, and indeed the vast majority ($\sim
        90\%$) are very radio weak. The remainder, however, display a
        wide range of radio power, from $\sim 10^{21-28} \, \whz$.  In
        this review I discuss recent radio imaging surveys that
        suggest there are three main classes of NLSy1, which cannot be
        easily distinguished by the standard radio-loudness parameter
        alone: (i) radio-weak NLSy1s without jets, (ii) mildly
        radio-loud NLSy1s that are a mixture of star-forming and
        jet-dominant AGN, and (iii) very radio loud NLSy1s with
        extreme properties similar to powerful jet-dominated blazars.
        I present updated kinematics information from the MOJAVE
        survey on six of the latter sources (all detected in
        gamma-rays by {\it Fermi)}, indicating high bulk Lorentz
        factors and small viewing angles in three cases.  Studies with
        the JVLA have shown that the jets of radio loud
        NLSy1s are likely lower-power versions of classical radio
        galaxies, with typical lengths of less than 10 kpc, although
        two very radio-loud NLSy1s have de-projected sizes of several
        hundred kpc. I discuss the challenges of reconciling the
        heterogeneous radio properties of NLSy1s with their strict
        optical line criteria, and near-term prospects for the
        discovery of larger numbers of radio-loud NLSy1s.

}

\FullConference{Revisiting narrow-line Seyfert 1 galaxies and their place in the Universe - NLS1 Padova\\
                9-13 April 2018 \\
                Padova Botanical Garden, Italy}
\begin{document}


\section{Background}
A longstanding challenge in understanding the AGN phenomenon is being
able to use the wealth of information obtained from optical
spectroscopy to predict and interpret properties at other wavebands.
The current list of known AGN is heavily dominated by
optically-selected objects, due to higher achievable \'etendue of
optical surveys compared to those in the radio, IR, X-ray, or
$\gamma$-ray, and more efficient spectroscopic followup studies. This
has historically led to the introduction of many AGN classes such as
narrow-line Seyfert 1 galaxies (NLSy1) that are purely
optically-based. As is frequently the case in astronomy, surveys of a
particular optical class at other wavelengths can reveal large
surprises, and NLSy1s are no exception in this regard.

After the NLSy1 classification criteria were formalized by
\cite{1985ApJ...297..166O} and \cite{1989ApJ...342..224G} in the
1980s, large X-ray surveys in the subsequent decade
\cite{1999AA...349..389V} revealed that NLSy1s show steep soft X-ray
continua, and are highly variable in the X-ray regime
\cite{1996AA...305...53B,2000NewAR..44..419H}. Discoveries of NLSy1
grew significantly during the 1990s, as follow-up spectroscopy
targeted X-ray sources displaying these properties (Table 1).

In the radio band, the situation proved to be quite different, as
hardly any of the approximately 200 known NLSy1s in the year 2000 were
known to have associated radio emission. It was only in 2007 that the
first cross-correlations of the FIRST VLA 1.4 GHz survey with SDSS
revealed $\sim 80$ radio loud NLSy1s \cite{2006AJ....131.1948W,
  2006ApJS..166..128Z}. In carefully selected samples
\cite{2006AJ....132..531K} the radio loud fraction was approximately
3-7\%, only half as large as the fraction seen in broad-lined AGN. The
low NLSy1 detection rates reflected both the limited sensitivity of
the available large radio surveys (e.g., NVSS and FIRST) and the
relatively low luminosities of NLSy1s in the radio. As for
$\gamma$-rays, the current situation echoes that of the radio band in
the early 2000s, with only a very small number of detections by the
{\it Fermi} LAT instrument, despite its continuous all-sky coverage
since 2008.  Although TeV $\gamma$-ray searches have been made
with air-shower Cherenkov telescopes
\cite{2014AA...564A...9H,2017AIPC.1792e0001B}, no significant NLSy1
detections have been reported to date.

\section{Why study NLSy1s in the radio?}
Given their small numbers and generally weak, hard to study radio
emission, it is perhaps surprising to see a large increase in the
number of NLSy1 radio studies in the last decade. Community interest
in this area has been driven by three main factors:

{\bf Imaging capabilities:} The radio band offers an unparalleled
ability to penetrate the dense gas in the host galaxy nucleus and
image the AGN emission in high detail. As described in
\S\ref{VLA}, the primary instruments in this regard are the JVLA and
VLBA, which offer kiloparsec- and pc-scale resolution imaging,
respectively, as well as full polarization capability. Both of these
instruments have received upgrades since 2010 that significantly
improved their flux sensitivity.  With the commissioning of ALMA in
2010, we have an additional powerful tool for studying star
formation and molecular gas at high resolution in low-luminosity AGN (LLAGN)
\cite{2018MNRAS.475.3004S,2017ApJ...845...50N,2016arXiv161206488S}.

{\bf Radio Loudness and the AGN Fundamental Plane:} NLSy1s appear to
defy the current paradigm that powerful jetted outflows are found
predominantly in elliptical galaxies with central black hole masses in
excess of $10^8 \Msun$. Although not many NLSy1 host galaxies have
been imaged to date, they appear (like normal Seyferts) to be mainly
spirals, but with a higher tendency towards bars
\cite{2007ApJS..169....1O}, pseudo-bulges \cite{2011MNRAS.417.2721O},
and denser environments \cite{2017AA...606A...9J}. NLSy1s also lie at
the extreme end of the Eigenvector 1 distribution (which relates Fe II
and H$\beta$ line widths with the X-ray continuum strength
\cite{2000ApJ...536L...5S}), which is likely a result of their high
Eddington ratios. Several AGN studies have shown a strong correlation
between the radio loudness parameter $R$\footnote{$R =
  S_\mathrm{radio} / S_\mathrm{optical}$, where $S$ are the flux densities at 5
  GHz and 4400 \AA, respectively} and black hole mass
\cite{2000ApJ...543L.111L,2003MNRAS.340.1095D,2004MNRAS.353L..45M}, as
well as an anti-correlation between $R$ and accretion rate.  Given
their low black hole masses and high accretion rates, NLSy1s would not
be expected to launch jets, and thus should be radio-quiet.  This
prediction appears to be borne out in low-fraction of radio-detected
NLSy1s, however, the general paradigm has difficulty accounting for
the sizable number of NLSy1s that are very radio loud (Table 1).

{\bf Jetted outflows in NLSy1s:} Studies with the VLA and MERLIN in the
1980s revealed that a large fraction of LLAGN generate
radio-synchrotron emitting jets
\cite{1984ApJ...285..439U,1986MNRAS.219..387U}. Although LLAGN jets
have much lower kinetic power than those of classical radio galaxies,
some are capable of escaping the confines of their host galaxy and
forming large isotropically-emitting radio lobes
\cite{1993ApJ...419..553B, 2018MNRAS.476.3478B}. Since they were not
expected to launch jets, only two NLSy1 radio imaging studies were
carried out prior to 2010 \cite{1995AJ....109...81U,
  2000NewAR..44..527M}, with the results of the latter only partially
presented in the proceedings of the 1999 Bad Honnef NLSy1 conference.
Both studies indicated that the radio emission was compact ($< 300$
pc; similar to Seyferts), although a subsequent larger study revealed
some NLSy1s with higher radio luminosities, $\simeq 10^{24} \whz$, and
evidence of variable radio emission \cite{2000NewAR..44..527M}. As
discussed in \S{\ref{VLA}}, direct imaging evidence of jetted outflows
in NLSy1s has been obtained only recently 
\cite{2012ApJ...760...41D,2015ApJ...800L...8R,2018arXiv180103519B}.
These images clearly show that NLSy1s are capable of generating
`triple' lobe-core-lobe radio jet structures reminiscent of
lower-power versions of classical radio galaxies.

\begin{table}
\begin{center}
{\label{table1} Table 1: Discovery History of NLSy1 Galaxies}  

\begin{tabular}{||l | c c c c||}
\hline
& & \bf{Year}& &\\
      \hline
\bf{Category} & 
\bf {1990}  &
\bf{2000} &
\bf{2010} &
\bf{2018} \\
      \hline
All &50 &205 &2049 &11472  \\
RL ($10 < R < 100$) &1 &2 &54 &234 \\
VRL ($R > 100$) &0 &1 &29 &144 \\
GeV $\gamma$-ray &0 &0 &4 &18 \\
TeV $\gamma$-ray &0 &0 &0 &0 \\
\hline
\end{tabular}

\end{center}
{\small {\bf Notes:} the columns indicate the number of known NLSy1s in each category as of the year indicated. RL = radio loud, VRL = very radio loud.}
\end{table}
\section{Radio Emission From AGN}
The radio loudness parameter $R$ figures prominently in many NLSy1
studies, as it is a readily observable quantity that can be tabulated
in large surveys. Its use as a jet/non-jet discriminator can be
problematic, however, in the case of NLSy1s since their jet powers tend
to be weak, and in most instances, may barely dominate other types of
radio emission from the host galaxy. For any given AGN, the total
radio output is a combination of (a) any nuclear (< 0.1 kpc) or (b)
larger scale (> 1 kpc) jet emission, (c) any emission from a
corona/wind on scales < 0.1 kpc, and (d) host galaxy star-formation
emission, in the form of supernovae and their remnants. High
sensitivity radio studies of nearby optically-selected AGN have shown
that at least one of these components is usually present, with truly
radio-silent AGN being extremely rare \cite{2016ApJ...831..168K}.

It can be very difficult or impossible to distinguish between these
radio-emitting components in AGN, especially when high resolution
imaging is not available. The few available studies with high quality
data have shown there is a great deal of variance in the relative
strengths of these components for individual AGN, particularly among
low-luminosity Seyferts
\cite{2013ApJ...768...37C,2006AJ....132..546G,1993ApJ...419..553B}.
Because radio emission from AGN jets can attain very high luminosities
($\sim 10^{29} \whz$; \cite{2017ApJ...842...87M}) whereas the most
powerful known starburst radio luminosities are only $\sim 10^{24}
\whz$, the launching of a jet can potentially increase the radio
loudness parameter $R$ of an AGN by several orders of magnitude. Early
studies of radio emission in bright optically selected quasar samples
\cite{1980AA....88L..12S,1989AJ.....98.1195K} showed indications of a
bi-modality in the radio loudness distribution at approximately $R
\simeq 10$, however, this result has been the subject of intense
debate involving selection effects and potential sources of error in
the $R$ parameter (see discussion by
\cite{2016ApJ...831..168K} and \cite{2003ApJ...583..145T,2001ApJ...555..650H}).
The latter include the effects of obscuration and host galaxy
contamination in the optical, uncertainties in k-correcting the $R$
value to the AGN rest frame due to incomplete spectral index
information, variability in the (usually non-simultaneously obtained)
radio and optical fluxes, and the influence of relativistic beaming on
the jet emission. More contemporary studies have found a continuous
distribution of $R$ in the AGN population \cite{2009AJ....137...42R,
  2011ApJ...743..104S}.  Indeed, the distributions of $R$, peak radio
flux density and luminosity for nearby quasars are all characterized
by a single peak that contains a mixture of star-forming and
jet-dominated AGN, plus an extended tail comprised of powerful jetted
AGN \cite{2016ApJ...831..168K}. The boundary at which star-forming AGN
begin to dominate the luminosity function is approximately $10^{23}
\whz$ at 5 GHz.

The observed 5 GHz radio luminosities of NLSy1s (given the current
large survey limits of a few mJy) span a very wide range, from $\sim
10^{21} \whz$ to $10^{28} \whz$ \cite{2018arXiv180103519B}. The median
1.4 GHz luminosity in the large NLSy1 sample of
\cite{2017ApJS..229...39R} is $10^{24} \whz$, right near the
star-forming/jet dominant boundary, and this is only for the 5\% of
NLSy1s that were detected in the 1.4 GHz FIRST survey. It is clear,
therefore, that the radio loudness parameter cannot be used as a
reliable jet indicator for the majority of NLSy1s.  In particular, the
oft-used $R= 10$ radio-loud/radio-quiet division, which was based on
the original \cite{1989AJ.....98.1195K} study, appears to fail when
applied to large samples of AGN and NLSy1s. The $R$ distribution for
the large compilation of NLSy1s in \cite{2017ApJS..229...39R} shows a
peak at $R=10$, with no evidence of bi-modality among the
radio-detected objects (the authors do not include any $R$ upper limit
values for NLSy1s without known radio counterparts). Also, there are
no clear changes seen in radio morphology \cite{2017MNRAS.466..921C}
or optical variability \cite{2017ApJS..229...39R} at $R\simeq 10$,
where the influences of a jet should supposedly be visible. It appears
clear, therefore, that in all but the most radio loud NLSy1s (i.e.,
those with R $\gtrsim$ 100), other indicators must be used as proof of
a jet, such as flux variability, the detection of compact radio
emission in VLA or VLBI images, or imaging of radio lobe emission
located outside the confines of the host galaxy.

\section{\label{VLA}Kiloparsec-scale Radio Properties of NLSy1s}
The general consensus on the radio structures of NLSy1s has been that
they are universally steep spectrum and compact, with sizes less than
300 pc. The latter number reflects the fact that prior to 2014, all
but 6 NLSy1s were apparently unresolved with the VLA, which can achieve
a typical resolution of $\sim 1''$ at frequencies of a few GHz.  Apart
from the small targeted imaging studies of \cite{1995AJ....109...81U}
and \cite{2000NewAR..44..527M}, which contained 7 and 24 sources,
respectively, our knowledge of NLSy1 radio structure up to that time
was based almost entirely on the FIRST \cite{2015ApJ...801...26H} and NVSS
\cite{1998AJ....115.1693C} surveys. These surveys, carried out at 1.4
GHz, have excellent sky coverage, but relatively poor angular
resolution ($5''$ and $45''$, respectively), and limited sensitivity
($\gtrsim 1$ mJy for FIRST, and $\gtrsim 2.5$ mJy for NVSS) for
studying radio quiet NLSy1s, or even the weak, small-scale radio
emission associated with radio-loud objects. The surveys differ mainly
in  their choice of array configurations, with the more extended
B-configuration of FIRST offering better angular resolution than NVSS
(D-configuration) at the expense of limited sensitivity to diffuse lobe emission.

Both FIRST and NVSS were carried out prior to a massive upgrade of the
VLA that was completed in early 2013.  The Jansky VLA (JVLA)
represents a factor of 10 improvement in sensitivity (capable of
attaining a 1 $\sigma$ continuum rms level of $<3 \mu$Jy in one hour
of integration), and offers broadband receivers from 1--50 GHz with
the capability of simultaneous in-band spectral index measurements.
The JVLA upgrade has enabled several recent studies that have
significantly revised our view of NLSy1 radio properties.

The study of \cite{2015ApJ...800L...8R} obtained JVLA observations of
three radio loud NLSy1s with the goal of getting accurate positions
for VLBI follow-up observations, and serendipitously discovered triple
radio structures in all three. Most of the radio lobes are
edge-brightened, with morphologies reminiscent of FR-II radio
galaxies, although with much lower radio power and jet lengths of only
10 to 30 kpc. However, given the fact that the pc-scale jets in these
NLSy1s appear one-sided \cite{2015IAUS..313..139R}, with the counter-jet
likely de-boosted by beaming effects, the de-projected sizes could be
as large as 400 kpc (see \S~\ref{kinematics}).

A much larger JVLA study of 74 NLSy1s, spanning a wide range of
redshift and radio loudness parameter $R$, was undertaken by
\cite{2018arXiv180103519B}. Nearly 2/3 of the sample were resolved
with the JVLA in its highest resolution configuration at 5 GHz (resolution
$\sim 0.3''$). Interestingly, most of the sources with $R < 10$ show
faint extended emission around an unresolved core, whereas very few of
the flat-radio spectrum NLSy1s in the sample display extended emission.
This is in contrast to powerful flat-spectrum radio quasars, which
frequently show weak diffuse kpc-scale emission consistent with radio
lobe structures viewed end-on \cite{1985ApJ...294..158A,
  2010ApJ...710..764K}.  The radio cores of the flat-spectrum NLSy1s
have brightness temperature lower limits of $10^4$ K to $10^8$ K, and many
are detected with VLBI (\S{\ref{VLBI}}), making it unlikely that star
formation is the source of their radio emission.  The $R < 10$ and
steep spectrum NLSy1s, on the other hand, have much lower core
brightness temperatures, so star formation cannot be ruled out, at
least in those sources that do not clearly display one-sided or
two-sided jet structures.

A main result of these two studies is that kpc-scale radio emission is
now known to be common in NLSy1s, although in the vast majority of
cases the overall projected extent is less than 10 kpc (similar to
ordinary Seyfert galaxies). There are currently only 16 known examples
with radio-emitting regions larger than 20 kpc, with the largest (1H
0323+342) having a projected extent of 140 kpc
\cite{2008AA...490..583A}. This rare group of NLSy1s are all very
radio loud ($ R > 100$), and 2/3 of them have classical lobe-core-lobe
morphology. {\it Fermi} has so far detected $\gamma$-ray emission from
6 of these NLSy1s, and the relatively high core dominance in their
radio images would suggest that their core emission is likely Doppler
boosted by relativistic bulk motion of pc-scale jets viewed at small
angles to the line of sight.

\section{\label{VLBI}Parsec-scale Radio Properties of NLSy1s}
Very Long Baseline Interferometry (VLBI) is a very efficient tool for
detecting jet emission in AGN, since current antenna arrays such as
the Very Long Baseline Array (VLBA) lack the short baselines and
sensitivity to detect radio emission with brightness temperature below
$\sim 10^5$ K. The typical frequencies of VLBI observations (1.4 GHz to 86
GHz) can easily penetrate the dust and gas found in the host galaxy
nucleus, offering a clear view of compact jet emission at
sub-milliarcsecond resolution ($\lesssim 5$ pc for the most distant
known NLSy1s). With multi-frequency and multi-epoch observations, it
is possible to precisely identify the location of the central engine,
perform spectral ageing studies on the lobe emission, and determine
the rates of jet expansion and overall age of the radio jets
\cite{2003PASA...20...19M,1998AA...337...69O}.

Several VLBI studies have found that jet activity is common in LLAGN,
with their radio cores showing a wide distribution of brightness
temperatures, from $10^5$ K to $10^8$ K
\cite{2004AA...417..925M,2004ApJ...603...42A,2009ApJ...706L.260G,
  2010MNRAS.401.2599O}. These studies typically recover only 50\% to
90\% of the total VLA flux density, indicating that most LLAGN have
low-surface brightness, diffuse emission on scales up to a few hundred
milliarcseconds that is invisible to VLBI \cite{2010MNRAS.401.2599O,
  2011ApJ...731...68L,2011nlsg.confE..64Z}. In individual LLAGNs where
the radio emission has been studied on different spatial scales (e.g.,
Mrk 6 \cite{2006ApJ...652..177K}, Mrk 1239
\cite{2015PASJ...67...15D}, NGC 6764 \cite{2010ApJ...723..580K}, NGC
4051 \cite{2009ApJ...706L.260G}, NGC 1068
\cite{2004ApJ...613..794G}, and NGC 2639 \cite{2010MNRAS.404.1966X}), the
radio morphology is often complex and distorted, and it is not clear
what fraction of the emission is from poorly collimated jets, winds
and/or star formation.

To date only $\sim 40$ NLSy1s have been observed with VLBI, mainly due
to the fact that their weak radio fluxes require time-consuming
phase-referencing observations, where atmospheric corrections to the
measured interferometric visibilities are determined by frequent
interleaved scans of a nearby bright calibrator source. Most of the
observations have thus focused on high $R$ sources, with the study of
7 NLSy1s by \cite{2013ApJ...765...69D} being the only one so far to
probe the radio-weak ($R < 10$) regime. The latter authors found that
both high- and low-$R$ sources can have core brightness temperatures
above $10^7$ K, and can display one-sided or two-sided jet morphology.
One-sided morphology was seen in 13 of 15 high $R$ jets by
\cite{2015IAUS..313..139R}, and in 7 of 14 very high $R$ jets by
\cite{2015ApJS..221....3G}, suggesting that the pc-scale emission of
highly radio loud NLSy1s is affected by Doppler boosting.  A main
difference in the pc-scale properties of NLSy1s with respect to other
LLAGN is the finding of much higher brightness temperatures (up to
$10^{11}$ K) in a small cohort that have been detected at GeV energies
by the {\it{Fermi}} LAT instrument
\cite{2012arXiv1205.0402O,2015ApJS..221....3G, MOJAVE_XIII}.

\begin{figure}
\centering
\includegraphics[angle=0,width=0.9\textwidth]{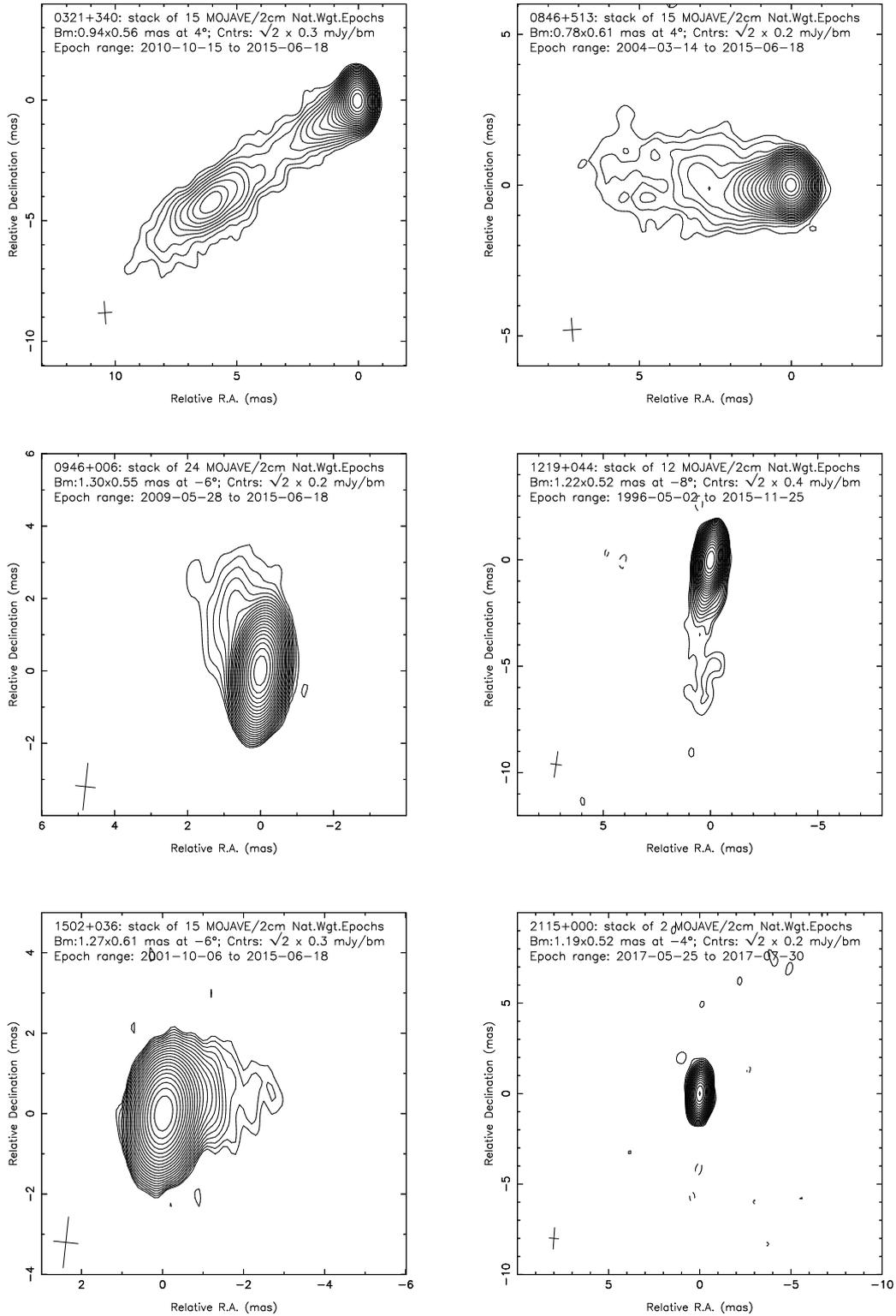}
  \caption{\label{VLBAimages} \small Stacked epoch 15 GHz VLBA images of six $\gamma$-ray loud NLSy1s from the MOJAVE program. 
}
\end{figure}

\section{\label{kinematics}Jet Kinematics of $\gamma$-ray NLSy1s}
The six brightest known NLSy1s in terms of radio flux density have
been the subject of regular monitoring with the VLBA at 15 GHz as part
of the MOJAVE program \cite{MOJAVE_V}. This program focuses on a
flux-limited sample of the brightest radio-loud AGN in the northern
sky, as well as radio-bright AGN detected by {\it{Fermi}}.  All of the
several hundred targeted AGN have correlated VLBA flux densities above
0.1 Jy at 15 GHz, and 96\% are blazars.  Doppler beaming can
significantly increase the observed radio flux by factors of up to
$\sim 10^5$ for highly relativistic jets viewed at small angles to the
line of sight \cite{1994ApJ...430..467V}, so the only non-blazars in
high frequency flux-density limited radio AGN surveys tend to be
nearby ($z < 0.1$) radio galaxies \cite{MOJAVE_I}.  Because
$\gamma$-ray emission in AGN is believed to originate near the bases
of powerful relativistic jets, the {{\it Fermi}} LAT AGN catalog
suffers from the same Doppler bias, and is also heavily dominated by
blazars \cite{3LAC}.  Approximately 3/4 of the radio-flux-limited
MOJAVE AGN were detected by {\it Fermi} in its first four years of
operation \cite{2015ApJ...810L...9L}.

The presence of six NLSy1s (all with $R> 1000$) in the MOJAVE survey
(Figure~\ref{VLBAimages}) is therefore significant, particularly more so
since all six are detected by {\it Fermi}. This is strongly indicative
of blazar-like properties.  The MOJAVE program
\cite{MOJAVE_XIII,Listerinprep} has obtained sufficient VLBA epochs to
analyze the jet kinematics of five of these $\gamma$-ray loud NLSy1s;
the results are listed in Table 2.  Three jets are highly
superluminal, while PMN J2118+0013 shows only sub-luminal motion over
a 5 year monitoring period, and PKS J1222+0413 shows no measurable
motion over a 14 year period.

\begin{figure}
\centering

\includegraphics[angle=270,width=0.9\textwidth]{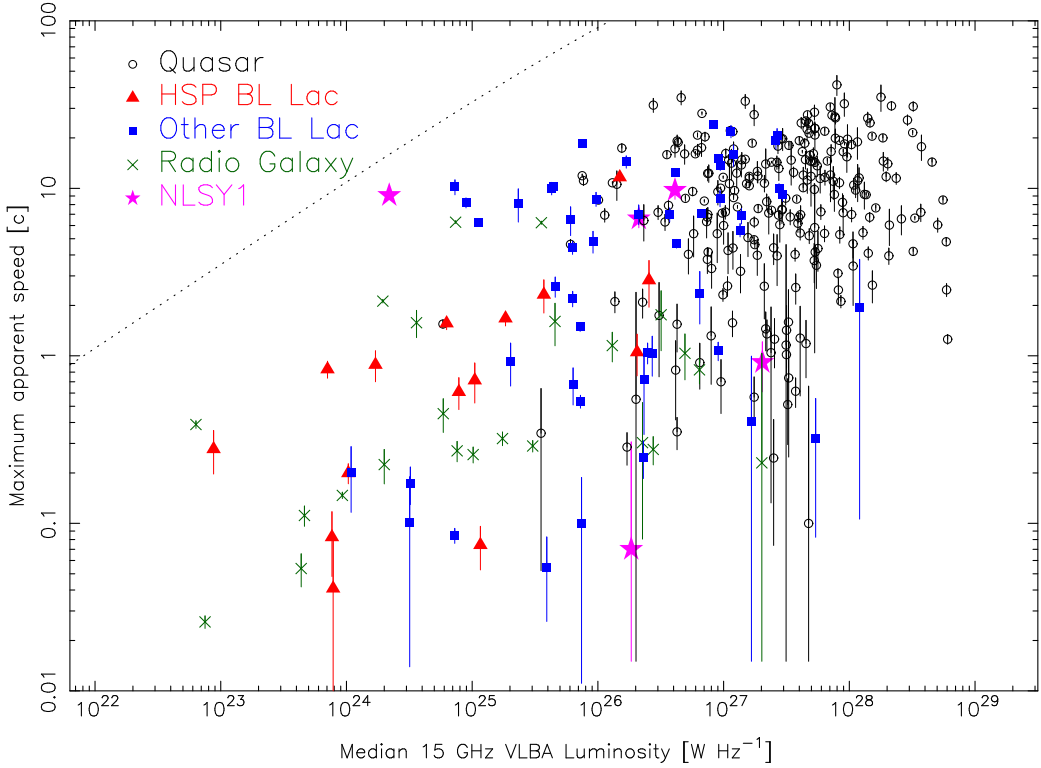}
  \caption{\label{betalum} \small Plot of maximum measured jet
    speed versus median 15 GHz VLBA radio luminosity for 334 jets in
    the MOJAVE sample from \cite{MOJAVE_XIII} and
    \cite{Listerinprep}. The dotted line indicates the upper limit of
    the survey.
}

\end{figure}

Figure~\ref{betalum} shows the maximum jet speed plotted against
median VLBA 15 GHz luminosity for 334 AGN in the MOJAVE sample. As
discussed by \cite{LM97}, the upper envelope to this distribution is a
result of the rarity of AGN jets with bulk Lorentz factors above $\sim
30$, and an intrinsic correlation between the true jet speed and
un-beamed radio synchrotron luminosity. Two of the $\gamma$-ray NLSy1s
(SBS 0846+513 and PMN J0948+0022) lie at the low end of the quasar
luminosity distribution, but have similar radio luminosities and
speeds to low-synchrotron peaked BL Lac objects. It is worth noting
the location of 1H 0323+342 at the extreme upper left edge of the
distribution.  This NLSy1 displays blazar-like variability at all
wavelengths \cite{2014ApJ...789..143P}, has appreciably high linear
polarization in its pc-scale jet \cite{MOJAVE_XV}, and appears to have
an atypically large apparent speed (9.1 c) for its radio luminosity
($2 \times 10^{24} \whz$). This could be indicative of a very high
intrinsic jet speed and/or high Doppler boosting factor (see, e.g.,
\cite{2007ApJ...658..232C}).  The lack of any appreciably measurable
jet speed in PKS 1502+036 may simply reflect that it has not ejected
any bright features during the five years it was monitored by the
MOJAVE program. In fact, there are 67 AGN in the MOJAVE sample which
do not have any $>3\sigma$ speed measurements \cite{Listerinprep}. In
the case of PMN J2118+0013, the VLBA has not yet revealed any
detectable emission downstream of its unresolved radio core\footnote{http://www.astro.purdue.edu/MOJAVE}.

The high superluminal speeds measured for three of these NLSy1s set
important constraints on their jet Lorentz factors ($\Gamma >
v_\mathrm{app} / c$) and jet viewing angles ($\theta < 2
\arctan(c/v_\mathrm{app})$). As can be seen in Table 1, the upper
limits of $\theta \lesssim 15^\circ$ imply de-projected sizes of $\gtrsim
600$ kpc for 1H 0323+342 and PMN J0948+0022, which, unless they
undergo significant jet bending from pc- to kpc-scales, would put them
towards the upper end of the size distribution for powerful radio
galaxies \cite{2017MNRAS.469.4083S}.  This sets them well apart from
the other radio-detected NLSy1s, which tend to have sizes below $\sim
10$ kpc.

\begin{table}
\begin{center}
{\label{table} Table 2: Kinematics of $\gamma$-ray Detected NLSy1 Jets}   
\begin{tabular}{|| l | c c c c c||}
\hline

{Name} & 
 {Redshift}  &
{$\beta_\mathrm{app}$} [c]&
{$\theta_\mathrm{max}$} [$^\circ$]&
{Size} [kpc] &
Refs. 
 \\
{(1)} & {(2)} & {(3)} & {(4)} &  
{(5)} & {(6)}
 \\
\hline
1H 0323+342	&0.061	&$9.1	\pm 0.3$&13	&140  	&\cite{MOJAVE_XIII},\cite{2008AA...490..583A},\cite{2009ApJ...707L.142A}\\
SBS 0846+513	&0.585	&$6.6	\pm 0.8$&17	&$<3$	&\cite{Listerinprep},\cite{2018arXiv180103519B},\cite{2012MNRAS.426..317D}\\
PMN J0948+0022	&0.5838	&$9.7	\pm 1.1$&12	&130	&\cite{Listerinprep},\cite{2012ApJ...760...41D}, \cite{2009ApJ...699..976A}\\
PKS J1222+0413 	&0.966	&$0.9	 \pm 0.3$&... 	&80	&\cite{Listerinprep},\cite{2010ApJ...710..764K},\cite{2015MNRAS.454L..16Y}\\
PKS 1502+036	&0.4083	&$0.07	 \pm 0.2$&...	&10	&\cite{Listerinprep},\cite{2018arXiv180103519B},\cite{2009ApJ...707L.142A}\\
PMN J2118+0013	&0.463	&...	         &...	&$<1.5$ &\cite{Listerinprep},\cite{2007ApJS..171...61H}, \cite{1FGL}\\
\hline
\end{tabular}
\end{center}
{\small Columns are as follows: (1) source name; (2) redshift; (3) maximum apparent jet feature speed, in units of the speed of light; (4) maximum jet viewing angle in degrees; (5) largest projected diameter of radio emission in kpc; (6) reference for apparent speed, kpc-scale radio image, and $\gamma$-ray discovery. 
}
\end{table}

\section{Summary}

In the last decade we have seen a flurry of investigations into the
radio properties of NLSy1s, with the number of known radio-loud
objects growing from just 3 to nearly 400. These studies have revealed
an unprecedented look into the jet activity of these AGN, whose low
black hole masses and high accretion rates should strongly disfavor
them from launching jets, according to the current AGN paradigm. On
the basis of their radio emission, NLSy1s appear to be divided into
three broad classes. The vast majority (>90\%) of NLSy1s are optically
non-variable \cite{2000PhDT.......105F} and very radio weak, being
undetectable above 1 mJy and having radio loudness parameters $R < 1$
\cite{2017ApJS..229...39R}.  Any radio emission in these sources is
likely to be from star formation or weakly collimated winds generated
by the accretion disk \cite{2010MNRAS.403.1246S,2015MNRAS.451.1795C}.
The second class contains what are typically termed radio-quiet or
mildly radio-loud objects, which have $1 < R < 100$, and are likely a
mixture of star-forming dominated and jet-dominated sources. Evidence
for the latter comes from the high fraction of NLSy1 that have
blazar-like colors in the IR \cite{2015MNRAS.451.1795C,
  2015AA...573A..76J} and X-ray variability that is much more
prevalent than in regular Seyferts
\cite{1996AA...305...53B,2006Natur.444..730M}. Their jets appear to be
lower-power versions of classical radio galaxies, with lengths less
than 10 kpc.  The rarest category, comprising less than 1\% of NLSy1s,
have powerful relativistic jets and display many of the
characteristics of flat-spectrum blazars, including high polarization
and multi-wavelength variability, and jet sizes up a to few hundred
kpc.  They can reach radio loudness values in excess of $R = 10^3$,
and 18 have been detected so far in $\gamma$-rays by
{\it Fermi}.

It remains to be determined why such an apparently heterogeneous
collection of AGNs can all meet the rigorous optical spectral line
criteria of the NLSy1 class. It has been speculated that orientation
effects of (a) geometrically thin broad line region \cite{2008MNRAS.386L..15D},(b) radiation pressure pushing the broad
line region to larger distances from the AGN
\cite{2008ApJ...678..693M}, and (c) relativistic beaming of optical
jet continuum \cite{2009MNRAS.396L.105G} all may affect observed line
strengths, but quantitative statistical assessment of these scenarios
awaits data on larger samples of radio-loud NLSy1s. In particular, it
will be important to conduct more VLBI studies on all types of LLAGN
that span a broad range of H$\beta$ line width, to avoid possible
selection biases in the NLSy1 class \cite{2012nsgq.confE..17V}.  The
VLA sky survey (VLASS), currently underway, will provide $2.5 ''$
resolution images, accurate positions, and flux densities from 2 to 5
GHz for over 5,000,000 radio sources in the entire sky north of
declination $-40^\circ$, down to limiting flux densities of 0.1 to 0.3
mJy. The combination of VLASS with large ongoing spectroscopic surveys
such as SDSS-BOSS \cite{2013AJ....145...10D} and LAMOST
\cite{2018AJ....155..189D} should result in many new NLSy1 candidates.
Followup near-IR spectroscopy longward of 10,000 $\angstrom$ will also
be fruitful for discovering NLSy1 past $z \simeq 0.9$, where the
$H\beta$ line is redshifted out of the optical range
(e.g.,\cite{2017FrASS...4....8B}).

\section*{Acknowledgments}

This conference has been organized with the support of the Department
of Physics and Astronomy ``Galileo Galilei'', the University of
Padova, the National Institute of Astrophysics INAF, the Padova
Planetarium, and the RadioNet consortium.  RadioNet has received
funding from the European Union's Horizon 2020 research and innovation
programme under grant agreement No~730562.  The MOJAVE program is
supported under NASA-Fermi grants NNX15AU76G and NNX12A087G.  The
National Radio Astronomy Observatory is a facility of the National
Science Foundation operated under cooperative agreement by Associated
Universities, Inc.

\bibliographystyle{JHEP}
\bibliography{lister}


\end{document}